\documentclass[12pt,preprint]{aastex}
\DeclareTextSymbol{\degre}{T1}{6}
\DeclareTextSymbol{\degre}{OT1}{23}
\usepackage{wasysym}

\slugcomment{Submitted to The Astrophysical Journal Letters}

\shorttitle{Removal of Titan's Atmospheric Noble Gases by their Sequestration in Surface Clathrates}
\shortauthors{O. Mousis et al.}

\begin{document}


\title{Removal of Titan's Atmospheric Noble Gases by their Sequestration in Surface Clathrates\\
}


 \author{
Olivier~Mousis\altaffilmark{1,*},
Jonathan~I.~Lunine\altaffilmark{2},
Sylvain~Picaud\altaffilmark{1},
Daniel~Cordier\altaffilmark{3,4},
J.~Hunter Waite, Jr.\altaffilmark{5},
and Kathleen~E.~Mandt\altaffilmark{5}}

\altaffiltext{1}{Universit{\'e} de Franche-Comt{\'e}, Institut UTINAM, CNRS/INSU, UMR 6213, Observatoire des Sciences de l'Univers de Besancon, France}

\altaffiltext{*}{Corresponding author E-mail address: olivier.mousis@obs-besancon.fr}

\altaffiltext{2}{Dipartimento di Fisica, Universit{\`a} degli Studi di Roma ``Tor Vergata'', Rome, Italy}
 
\altaffiltext{3}{Universit{\'e} de Rennes 1, Institut de Physique de Rennes, CNRS, UMR 6251, France}

\altaffiltext{4}{Ecole Nationale Sup{\'e}rieure de Chimie de Rennes, France}

\altaffiltext{5}{Space Science and Engineering Division, Southwest Research Institute, San Antonio, TX 78228, USA}

\begin{abstract}
A striking feature of the atmosphere of Titan is that no heavy noble gases other than argon were detected by the Gas Chromatograph Mass Spectrometer (GCMS) aboard the Huygens probe during its descent to Titan's surface in January 2005. Here we provide an explanation of the mysterious absence or rarity of these noble gases in Titan's atmosphere: the thermodynamic conditions prevailing at the surface-atmosphere interface of the satellite allow the formation of multiple guest clathrates that preferentially store some species, including all heavy noble gases, over others. The clean water ice needed for formation of these clathrates could be delivered by successive episodes of cryovolcanic lavas that have been hypothesized to regularly cover the surface of Titan. The formation of clathrates in the porous lavas and their propensity for trapping Ar, Kr and Xe would progressively remove these species from the atmosphere of Titan over its history. In some circumstances, a global clathrate crust with an average thickness not exceeding a few meters could be sufficient on Titan for a complete removal of the heavy noble gases from the atmosphere.

\end{abstract}





\section{Introduction}
A striking feature of the atmosphere of Titan is that no heavy noble gases other than argon were detected by the Gas Chromatograph Mass Spectrometer (GCMS) aboard the Huygens probe during its descent to Titan's surface in January 2005 (Niemann et al. 2005, 2010). The detected argon includes primordial $^{36}$Ar, present in subsolar abundance in Titan's atmosphere ($^{36}$Ar/$^{14}$N is found to be about six orders of magnitude lower than the solar value), and the radiogenic isotope $^{40}$Ar, which is a decay product of $^{40}$K (Niemann et al. 2005). The other primordial noble gases $^{38}$Ar, Kr and Xe were not detected by the GCMS instrument, yielding upper limits of 10$^{-8}$ for their atmospheric mole fractions (Niemann et al. 2005, 2010).

In order to interpret this deficiency, it has been proposed that the atmospheric depletion of these species could be explained by their dissolution at ambient temperature in the hydrocarbon lakes and seas present on Titan's surface (Cordier et al. 2010). However, the fractions of argon and krypton that would dissolve in these liquids were found to be negligible (Cordier et al. 2010). Another interpretation of the lack of heavy noble gas abundances above 10 ppb in Titan's atmosphere is based on laboratory results suggesting that the haze present in Titan's atmosphere could efficiently trap argon, krypton and xenon during its formation (Jacovi \& Bar-Nun 2008). In this mechanism, the open structure of the small aerosol particles would allow the noble gas atoms to fill their pores. However, even if this trapping mechanism is effective, the erosion of sedimented aerosols might induce the progressive release of the trapped noble gases in Titan's atmosphere.

The presence of clathrate hydrates on Titan's surface has also been proposed to be the origin of the heavy noble gas deficiency measured in its atmosphere (Osegovic \& Max 2005). A series of theoretical investigations showed that the trapping efficiency of these ice structures is high enough to significantly decrease the atmospheric concentrations of xenon and krypton, but failed to explain the observed argon atmospheric deficiency (Thomas et al. 2007, 2008). It was then argued that Titan's building blocks were partly devolatilized within a subnebula around the forming Saturn, implying the loss of argon initially incorporated as pure condensate prior to satellite formation (Mousis et al. 2009). However, more recent work showed that argon could have been retrapped in primordial clathrates so as to remain abundant in the planetesimals incorporated in Titan (Mousis et al. 2010). Given the evidence from $^{40}$Ar that Titan is extensively outgassed (Niemann et al. 2010), it is difficult to explain the observed $^{36}$Ar rarity in the atmosphere in the absence of a sink over geologic time.

Here we show that trapping over time of atmospheric argon in clathrates could indeed provide such a sink. This conclusion is based on the use of recent intermolecular potential parameters in a statistical thermodynamic model describing clathrate 
composition. The fresh water ice needed for formation of these clathrates could be delivered by successive episodes of cryovolcanic lavas hypothesized to periodically cover the surface of Titan. The formation of clathrates in the porous cryo-lavas and their propensity for trapping Ar, Kr and Xe would progressively remove these species from the atmosphere of Titan over its history. A global clathrate crust with an average thickness not exceeding a few meters could be sufficient on Titan for a complete removal of the heavy noble gases from the atmosphere.

\section{The statistical-thermodynamic model}
\label{model}

To calculate the relative abundances of guest species incorporated in a clathrate from a coexisting gas of specified composition at given temperature and pressure, we use a model applying classical statistical mechanics that relates the macroscopic thermodynamic properties of clathrates to the molecular structure and interaction energies (van der Waals \& Platteuw 1959; Lunine \& Stevenson 1985). It is based on the original ideas of van der Waals and Platteeuw for clathrate formation, which assume that trapping of guest molecules into cages corresponds to the three-dimensional generalization of ideal localized adsorption. In this work, the availability of recent intermolecular potential parameters retrieved from experiments (Sloan \& Koh 2008; hereafter SK08) allows us to revise the relative abundances of guest species incorporated in a clathrate from a coexisting gas of specified composition at temperature and pressure relevant to Titan's surface conditions.
	
In this formalism, the fractional occupancy of a guest molecule $K$ for a given type $t$ ($t$~=~small or large) of cage can be written as

\begin{equation}
\label{occupation}
y_{K,t}=\frac{C_{K,t}P_K}{1+\sum_{J}C_{J,t}P_J} ,
\end{equation}

\noindent where the sum in the denominator includes all the species which are present in the initial gas phase. $C_{K,t}$ is the Langmuir constant of species $K$ in the cage of type $t$, and $P_K$ is the partial pressure of species $K$. This partial pressure is given by $P_K=x_K\times P$ (we assume that the sample behaves as an ideal gas), with $x_K$ the mole fraction of species $K$ in the initial gas, and $P$ the total atmospheric gas pressure, which is dominated by N$_2$. The Langmuir constant depends on the strength of the interaction between each guest species and each type of cage, and can be determined by integrating the molecular potential within the cavity as

\begin{equation}
\label{langmuir}
C_{K,t}=\frac{4\pi}{k_B
T}\int_{0}^{R_c}\exp\Big(-\frac{w_{K,t}(r)}{k_B T}\Big)r^2dr ,
\end{equation}

\noindent where $R_c$ represents the radius of the cavity assumed to be spherical, $k_B$ the Boltzmann constant, and $w_{K,t}(r)$ is the spherically averaged Kihara potential representing the interactions between the guest molecules $K$ and the H$_2$O molecules forming the surrounding cage $t$. This potential $w(r)$ can be written for a spherical guest molecule, as (McKoy \& Sinano\u{g}lu 1963)

\begin{eqnarray}
\label{pot_Kihara}
w(r) 	= 2z\epsilon\Big[\frac{\sigma^{12}}{R_c^{11}r}\Big(\delta^{10}(r)+\frac{a}{R_c}\delta^{11}(r)\Big)
- \frac{\sigma^6}{R_c^5r}\Big(\delta^4(r)+\frac{a}{R_c}\delta^5(r)\Big)\Big],
\end{eqnarray}

\noindent with

\begin{equation}
\delta^N(r)=\frac{1}{N}\Big[\Big(1-\frac{r}{R_c}-\frac{a}{R_c}\Big)^{-N}-\Big(1+\frac{r}{R_c}-\frac{a}{R_c}\Big)^{-N}\Big].
\end{equation}

\noindent In Eq. (\ref{pot_Kihara}), $z$ is the coordination number of the cell. This parameter depends on the structure of the clathrate (I or II) and on the type of the cage (small or large). The Kihara parameters $a$, $\sigma$ and $\epsilon$ for the molecule-water interactions, given in Table \ref{Kihara}, have been taken from the recent work of SK08 when available and from Parrish \& Prausnitz (1972) (hereafter PP72) for the remaining species.

Finally, the mole fraction $f_K$ of a guest molecule $K$ in a clathrate can be calculated with respect to the whole set of species considered in the system as

\begin{equation}
\label{abondance} f_K=\frac{b_s y_{K,s}+b_\ell y_{K,\ell}}{b_s \sum_J{y_{J,s}}+b_\ell \sum_J{y_{J,\ell}}},
\end{equation}

\noindent where $b_s$ and $b_l$ are the number of small and large cages per unit cell respectively, for the clathrate structure under consideration, and with ${\sum_{K}} f_{K}~=~1$. Values of $R_c$, $z$, $b_s$ and $b_l$ are taken from PP72.

\section{Results}

Table \ref{atm} gives the composition of Titan's atmosphere used in our calculations. We made the conservative assumption that all noble gases were initially present in the atmosphere of Titan, with Ar/N, Kr/N and Xe/N ratios assigned to be solar (Asplund et al. 2009). Because our composition calculations are only valid along the equilibrium curve of the clathrate of interest, they were performed at 165.8 K, i.e. the equilibrium temperature of a multiple guest clathrate formed from the atmosphere of Titan possessing a pressure of 1.46 bar at the surface (see Appendix). Composition calculations have been performed in the cases of structures I and II clathrates. We have also taken into account the influence of variation of the cage sizes on the mole fractions of guests encaged in clathrates by modifying the values of their initial radius by up to 3\%. Indeed, thermal expansion or contraction measured in the temperature range of 90--270 K have been found to significantly affect the composition of clathrates as well as their dissociation pressures that define the (P,T) loci of stability (Mousis et al. 2010; Belosludov et al. 2002).

\begin{figure}
\includegraphics[angle=-90,width=11cm]{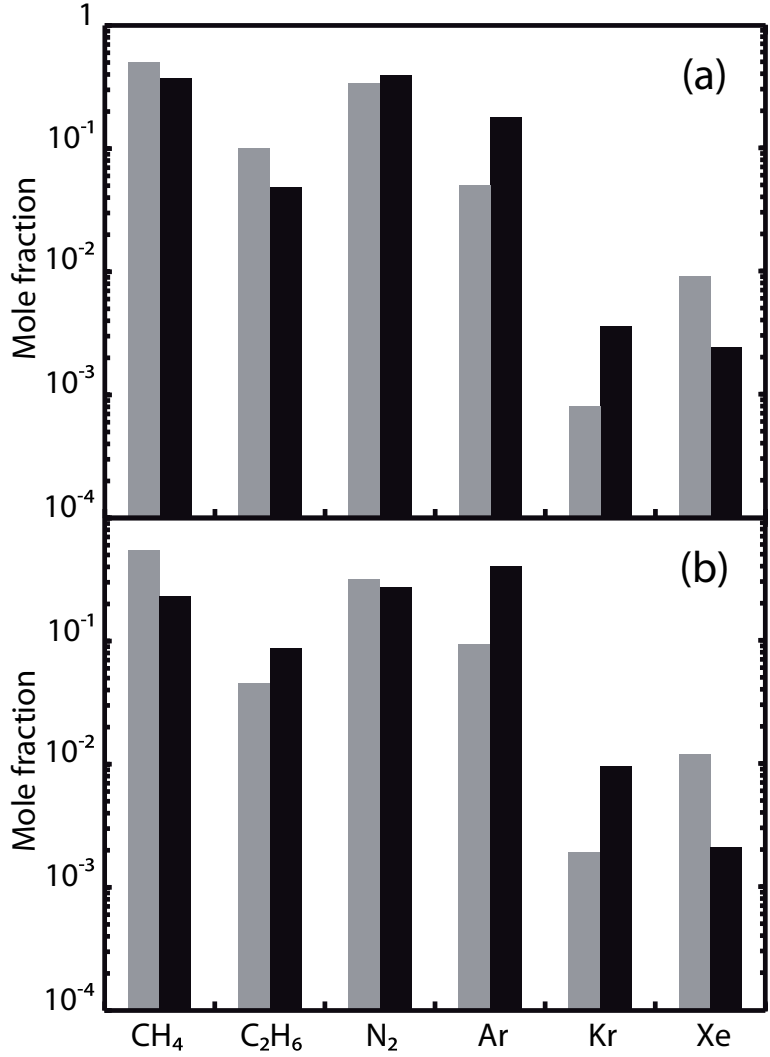}
\caption{Mole fraction $F_K$ of volatile $K$ encaged in multiple guest clathrate formed on Titan's surface at $T$ = 165.8 K and $P$ = 1.46 bar. Calculations are performed in the cases of (a) our nominal model (no cage variation) and (b) a 2\% contraction of the cages. Grey and dark bars correspond to structure I and structure II clathrates, respectively.} 
\label{comp1}
\end{figure}

Figure \ref{comp1} gives the composition of structures I and II clathrates expected to form on the surface of Titan from the nitrogen-dominated atmosphere in the cases of our nominal model (no cage variation) and a 2\% contraction of the cages (Belosludov et al. 2002). Assuming that noble gases are in solar abundances relative to elemental nitrogen (Asplund et al. 2009), the total number of Ar, Kr and Xe atoms that would exist in the atmosphere of Titan is order of 1.06~$\times$~10$^{43}$, 3.23~$\times$~10$^{39}$ and 2.02~$\times$~10$^{38}$, respectively. The equivalent volume of clathrate needed to fully trap argon, the most abundant noble gas, would be then about 2.29~x~10$^{15}$ m$^{\rm 3}$ in clathrates, irrespective of the considered structure\footnote{The unit cell of structure I consists of 8 cages for a unit volume of 1.73 $\times$ 10$^{-27}$ m$^{\rm 3}$ while the unit cell of structure II consists of 24 cages for a unit volume of 5.18 $\times$ 10$^{-27}$ m$^{\rm 3}$.}. On the other hand, since the mole fraction of Ar has been found to be about 5.01 $\times$ 10$^{-2}$ and 1.78 $\times$ 10$^{-1}$ in structures I and II clathrates of our nominal model, we then obtain a total clathrate volume of 4.55 $\times$ 10$^{16}$ and 1.28 $\times$ 10$^{16}$ m$^{\rm 3}$ needed to trap simultaneously the three noble gases in these two structures, respectively. In the case of a 2\% contraction of the cages, the mole fraction of Ar becomes 9.24 $\times$ 10$^{-2}$ and 4.02 $\times$ 10$^{-1}$ in structures I and II clathrates, corresponding to a total clathrate volume of 2.47 $\times$ 10$^{16}$ and 5.67 $\times$ 10$^{15}$ m$^{\rm 3}$ in these two structures, respectively. Interestingly, irrespective of the considered case, Ar remains the less efficiently trapped of the three considered noble gases (respectively to their initial atmospheric abundances) so its complete enclathration also implies the full trapping of krypton and xenon.

\begin{figure}
\resizebox{\hsize}{!}{\includegraphics[angle=90]{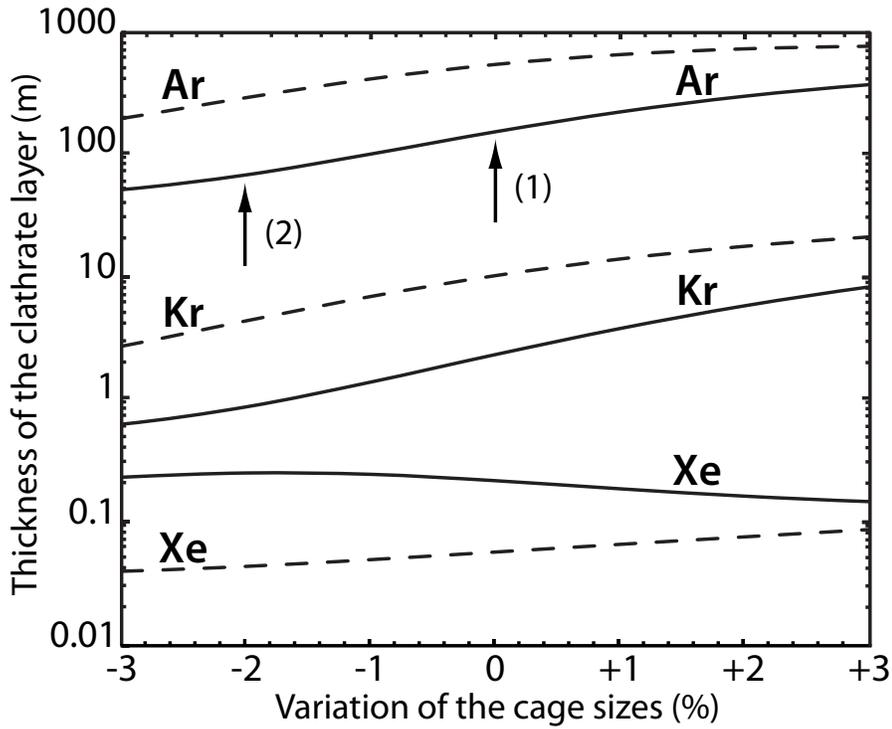}}
\caption{Thickness of structure I (dashed lines) and structure II (solid lines) clathrate layers needed to trap Ar, Kr or Xe as a function of the cage sizes. Vertical arrows give the value of the thickness of the structure II clathrate layer needed to simultaneously trap all noble gases in the cases of (1) our nominal model (no cage variation) and (2) a 2\% contraction of the cages. Calculations have been made at $T$ = 165.8 K and $P$ = 1.46 bar.} 
\label{thick}
\end{figure}

Figure \ref{thick} represents the thickness of the clathrate layer needed to trap the three noble gases as a function of the variation of the cage sizes in the cases of structures I and II clathrates. In the case of the nominal model, the total clathrate volume translates into a multiple guest clathrate layer with an equivalent thickness of about 549.6 and 154.7 m globally averaged in structures I and II, respectively. If one considers the case corresponding to a 2\% contraction of the clathrates cages, then the equivalent thickness of the clathrate layer would be reduced to about 298.0 and 68.6 m over the surface of Titan in structures I and II, respectively.\\
 
\section{Discussion}
\label{discussion}

Formation scenarios predict that Titan only accreted elemental nitrogen in the form of ammonia during its formation (Mousis et al. 2002, 2009), this latter being converted into molecular nitrogen via photolysis (Atreya et al. 1978) or shock chemistry (McKay et al. 1988) in the atmosphere. If 5\% of solar elemental nitrogen was accreted by Titan in the form of ammonia (Mousis et al. 2002), then the equivalent thickness of the clathrate layer would reduce to about 27.4 and 7.7 m in the nominal model and 14.8 and 3.4 m in the case of a 2\% contraction of the cages in structures I and II, respectively. Our model then suggests that formation scenarios favoring the accretion of Titan from NH$_3$-rich and N$_2$-poor planetesimals are preferable because the amount of clathrates needed to explain the noble gas atmospheric deficiency is much less important than in the case of satellite accretion from N$_2$-rich planetesimals.
  
It should be noted that, because i) the atmosphere of Titan is by far dominated by N$_2$ and ii) N$_2$ clathrate is of structure II (Lunine \& Stevenson 1985; SK08), we should expect the presence of structure II clathrates on the satellite's surface. In any case, future laboratory experiments aiming at investigating the formation of clathrates formed from a mixture analogous to that of Titan's atmosphere could help to disentangle between the two possible structures. In this context, assuming that Titan's clathrates are of structure II, the average thickness of the clathrate layer needed on Titan for the complete removal of the heavy noble gases from the atmosphere should then range {between a few tens of meters to a few hundreds of meters.} 

Our calculations imply that noble gas-rich clathrates either formed (i) when the surface of the satellite of Saturn was warmer than today (Thomas et al. 2007) and/or (ii) during more recent release events of hot and porous cryolava possibly associated with methane delivery towards the surface (Mousis \& Schmitt 2008; Tobie et al. 2006). In the first case, Titan's accretional heating would have resulted in an initially warm surface ($\sim$300--500 K), implying the melting of the ice (Kuramoto \& Matsui 1994; Lunine et al. 2010). With time, crustal freezing would have allowed formation of multiple guest clathrates at the surface/atmosphere interface (Tobie et al. 2006; Lunine et al. 2010). In the second case, a highly porous icy material should be generated from the decomposition of methane clathrates (Mousis \& Schmitt 2008) and the formation of multiple guest clathrates at the surface-atmosphere interface would take place during the cooling of the cryolava, which needs about 1 Earth yr to decrease to Titan's surface temperature (Lorenz 1996). In both cases, the clathrates' composition is assumed to remain fixed during the cooling of Titan's surface.

Because the volume expansion induced by clathrate formation ($\sim$20\% compared to that of water ice (Mousis \& Schmitt 2008)) may well cause its self-isolation by closing the pores of the ice that is not in direct contact with the atmosphere, the thickness of the clathrate layer formed during the satellite cooling should be very limited, unless events such as impacts {(Sekine et al. 2011)} expose clean ice. Cryovolcanic activity, whose features are visible in some regions of Titan (Wall et al. 2009), is probably required to create substantial amounts of porous water ice spread across the surface, in a way similar to basaltic lava flows on Earth. Long periods of active cryovolcanism on Titan (Tobie et al. 2006) should lead to the continuous build up of consecutive layers of cryolava (Mousis \& Schmitt 2008) and thus favor the formation of noble gas rich clathrate layers. The GCMS detection of $^{40}$Ar remains consistent with our model and implies that it is the most abundant argon isotope trapped in the clathrates layers. Its release into the atmosphere would then result from outgassing events (Waite et al. 2005) that occurred later than the removal of all the atmospheric noble gases. 

However, the formation of noble gas-rich clathrate layers from water ice delivered via cryovolcanism might meet two limitations. First, once in contact with the atmosphere, the cryolava flows might be too hot to directly form noble gas-rich clathrate layers. Second, cryolava flows could act as sources of noble gases by liberating them from underlying clathrates through melting. The volume
(temporarily) melted by the flow may well exceed the volume of the flow itself that is available for sequestering, resulting in net release of noble gas, instead of net sequestration. On the other hand, clathrate formation should remain possible during the rapid cooling of the cryolava. Indeed, the equilibrium pressure of a N$_2$-dominated clathrate (similar to our structure II nominal model) is about 10$^{-3}$ bars for a surface temperature of 94 K. With a value of $\sim$ 1.3 bars, the partial pressure of N$_2$ in the atmosphere greatly exceeds the equilibrium pressure of N$_2$-dominated clathrate and then favors the formation of this structure at the current surface temperature. As such, the newly formed clathrate layers could at least partially absorb the noble gases recently released into the atmosphere by hot cryolava flows.
\acknowledgements
We thank an anonymous Referee whose comments have helped us improve our manuscript.

\section{Appendix}

In the present study, the temperature dependence of the dissociation pressure of the multiple guest clathrate is determined from available experimental data and from a combination rule due to Lipenkov \& Istomin (2001). Thus, the dissociation pressure $P^{diss}_{mix}$ of a multiple guest clathrate is calculated from the dissociation pressure   $P^{diss}_{K}$ of a pure clathrate of guest species $K$ as 

\begin{equation}
P^{diss}_{mix} = \left [\sum_K \frac{x_K}{P^{diss}_{K}} \right ]^{-1},
\end{equation}

\noindent where $x_K$ is the molar fraction of species $K$ in the gas phase. The dissociation pressure $P^{diss}_{K}$  is derived from laboratory measurements and follows an Arrhenius law (Miller 1961):

\begin{equation}
log(P^{diss}_{K}) = A + \frac{B}{T},
\end{equation}

\noindent where $P^{diss}_{K}$ and $T$ are expressed in Pa and K, respectively. The constants $A$ and $B$ used in the present study have been fitted to the experimental data used by Thomas et al. (2007) and derived from Lunine \& Stevenson (1985) and SK08.

%
%

\clearpage
\begin{table}[h]
\centering \caption{Parameters for the Kihara potential}
\begin{tabular}{clccc}
\hline \hline
Molecule   & $\sigma$(\AA)& $ \epsilon/k_B$(K)& $a$(\AA) 	& Reference	\\
\hline
N$_2$			& 3.13512		& 127.426		& 0.3526	&	SK08\\
CH$_4$     		& 3.14393     	& 155.593     	& 0.3834 	&	SK08\\
C$_2$H$_6$       	& 3.24693     	& 188.181     	& 0.5651 	&	SK08\\
Ar              			& 2.9434     	& 170.50     	& 0.184 	&	PP72\\
Kr             			& 2.9739     	& 198.34     	& 0.230 	&	PP72\\
Xe              		& 3.32968     	& 193.708     	& 0.2357 	&	SK08\\
\hline
\end{tabular}
\tablecomments{$\sigma$ is the Lennard-Jones diameter, $\epsilon$ is the depth of the potential well, and $a$ is the radius of the impenetrable core.}
\label{Kihara}
\end{table}

\clearpage
\begin{table}[h]
\centering \caption{Assumed composition of Titan's atmosphere at the ground level}
\begin{tabular}{lcc}
\hline \hline
Species $K$			& Mole fraction $f_K$		& Reference  				\\
\hline
N$_2$			& 8.80 $\times$ 10$^{-1}$			& This work				\\
CH$_4$     		& 4.92 $\times$ 10$^{-2}$   		& Niemann et al. 2005 		\\
C$_2$H$_6$       	& 1.49 $\times$ 10$^{-5}$   		& Cordier et al. 2009   		\\
Ar              			& 7.07 $\times$ 10$^{-2}$   		& This work    				\\
Kr             			& 5.01 $\times$ 10$^{-5}$   		& This work    				\\
Xe              		& 4.89 $\times$ 10$^{-6}$   		& This work    				\\
\hline
\end{tabular}
\label{atm}
\end{table}

\clearpage

\clearpage

\end{document}